%% file: sample-sigconf-authordraft.tex
\documentclass[sigconf,screen,nonacm]{acmart}

\AtBeginDocument{%
  }

\begin{document}

\title{Make Any Collection Navigable: Methods for Constructing and Evaluating Hypergraph of Text}

\author{Dean E. Alvarez}
\affiliation{%
  \institution{University of Illinois Urbana-Champaign}
  \city{Urbana}
  \state{IL}
  \country{USA}}
\email{deana3@illinois.edu}

\author{ChengXiang Zhai}
\affiliation{%
  \institution{University of Illinois Urbana-Champaign}
  \city{Urbana}
  \state{IL}
  \country{USA}}
\email{czhai@illinois.edu}

\begin{abstract}
    One reason the Web is more useful than a simple collection of documents is that the structure created by hyperlinks enables flexible navigation from one web page to another. However, hyperlinks are typically created manually and cannot fully capture a corpus' implicit semantic structures. Is there a general way to make an arbitrary collection navigable? Recent work has formalized this problem generally as constructing a Hypergraph of Text (HoT), which provides a formal mathematical structure for supporting navigation and browsing. However, how to construct and evaluate a Hypergraph of Text remains a challenge. In this paper, we propose and study several methods for constructing a HoT. We also propose a novel quantitative metric, effort ratio, for evaluating the structural quality of a constructed HoT. Experimental results show that even simple TF-IDF baselines can match LLM-based methods on our proposed effort ratio metric. 
\end{abstract}

\maketitle

\input{sections/intro}
\input{sections/related_work}
\input{sections/HoT_Construction}
\input{sections/methods}
\input{sections/effort_ratio}
\input{sections/results}
\input{sections/conclusion}
\section*{Acknowledgment}
This work is supported by a grant from Talroo Inc. It is also supported in part by the National Science Foundation (NSF) under grant number 2433308.

\bibliographystyle{ACM-Reference-Format}
\bibliography{ref}

\end{document}

%% file: sections/intro.tex
\section{Introduction}

Despite the massive progress in the field of information retrieval over the past decades, discovering and navigating relevant content in a scattered collection remains a challenge. 
Although recent progress in Large Language Models (LLMs) has addressed this challenge to some extent (notably via direct support of question answering), concerns about their trustworthiness~\cite{Ji2022SurveyOH} and their lack of access to the most recent information limit their utility. How to enable users to have direct access to and discover the most relevant original information is still an open challenge. 

While search engines and LLM tools such as ChatGPT are useful to help people find relevant information, they require users to formulate an effective query or question, thus limiting their utility when there is a vocabulary gap. Moreover, they are less helpful helping users discover relevant but unknown information. 

Browsing has been proposed as a way to overcome these difficulties by enabling users to explore the information space without needing any prior knowledge on how to formulate queries, what vocabulary is used in the document collection, nor what sort of things are in the collection at all. Browsing allows for a greater degree of user agency and clearer information provenance than LLM QA systems, while allowing for more serendipity than traditional search. These features make it an attractive paradigm for tasks such as exploratory information retrieval or collection understanding.  

While browsing has long been of interest to many within the IR and Web communities (see,  e.g.,~\cite{bates1989design,bates2007browsing,marchionini1995information,rice2001accessing,white2009exploratory,aula2010does,chi2001using,singer2017we,clemm2024analysis}), it has seen relatively less research progress toward general technologies to support browsing compared to those to support search and question answering. A key difference between browsing and search or question answering is that while search and question answering have enjoyed unified mathematical frameworks and intuitive quantitative metrics, browsing has not. As a result, while many previous studies have argued for the need to support browsing (see, e.g.,~\cite{bates1989design,marchionini1995information, white2009exploratory}) and much work has been done on browsing, there is still a lack of a general tool to organize an arbitrary collection of text and enable navigation in the entire space; users today mostly only navigate through manually created links such as the many hyperlinks on the Web. 

Recently, Hypergraph of Text (HoT) ~\cite{HoT} was proposed as a mathematical structure to organize an arbitrary collection of text documents into a hypergraph semantic structure and make it navigable. The intuition is that a hyperedge provides a general way to define any semantic neighborhood, a key concept enabling semantic navigation. HoT can serve as a theoretical framework for building a general navigation system in the sense that it is possible to represent any browsable collection of documents as a HoT and any HoT can be visually browsed on a uniform interface.  Indeed, the Text Information Navigation Kit~\cite{TINK25} can be used to visualize any HoT and enable a user to navigate the information space represented by the HoT. 

However, how to construct a HoT and how to evaluate the utility of the browsing support provided by a HoT to a user remain open challenging questions that are not well studied; the original work~\cite{HoT} only introduced a single method which was evaluated qualitatively.   

In this paper, we seek to address this knowledge gap by studying how to tackle both challenges. Specifically, we study several approaches to HoT construction such as LLMs, embedding, and TF-IDF filtering. We address the evaluation challenge by proposing 
a novel metric that quantifies the utility of the HoT's hyperedges for browsing. Using the proposed evaluation method, we systematically compared multiple approaches of HoT construction on a Web dataset. Experimental results show that while various methods can create effective links, each involves distinct tradeoffs and LLM-based approaches do not always outperform simpler statistical approaches in terms of structural navigation.  

In sum, the contributions of this paper include: 
\begin{enumerate}
    \item We propose and study three methods (TF-IDF filtering, LLM topic extraction, and two step similarity-then-LLM) for solving the HoT construction problem.
    \item We propose a novel metric, which enables quantitative evaluation of HoT construction algorithms.
    \item  We conduct experiments to evaluate HoT construction algorithms and find that LLM-doc achieves the lowest effort ratio but at the cost of 38\% disconnected relevant pairs, and that TF-IDF-based All-Words is competitive on structure despite being orders of magnitude cheaper.
\end{enumerate}
Overall, our work demonstrates the great potential of developing general technologies for supporting browsing based on HoT construction. 

%% file: sections/related_work.tex
\section{Related Work}
\label{sec:related}
{\bf Information Extraction:} Information Extraction~\cite{yang2022survey} has been studied extensively for many decades with some recent work focused on extraction of entities and relations to construct a knowledge graph~\cite{zhong2023comprehensive}.
Traditional methods rely on supervised learning, notably the Conditional Random Field approaches~\cite{lafferty2001conditional,sutton2012introduction} and focus more on pre-defined entity and relation types. Later approaches explored how to perform information extraction with open entities and relations~\cite{niklaus2018survey}. 

Recently, LLMs have been leveraged for knowledge graph construction~\cite{xu2023large, pan2024unifying}. Although KG could be used to organize text data, its coverage of text content is limited to the entities and relations included in the KG; in contrast, the links generated for HoT construction cover the entire information space and enable a user to navigate into potentially any region in the information space. In terms of the information extracted, HoT construction aims to extract topical information, whereas Information Extraction is more focused on extracting entities and relations. The information extracted in HoT construction is more useful for supporting user navigation, whereas the extracted entities and relations tend to be more useful for constructing a knowledge graph to further support all kinds of downstream applications in NLP.  

{\bf Topic Models:} Probabilistic topic models~\cite{vayansky2020review,abdelrazek2023topic} are general techniques to discover and analyze topics covered in the text data. The extracted topics are usually represented as word distributions. There are many applications of the discovered topics and their coverage in the documents, especially opinion analysis, spatiotemporal topic trend analysis, and document representation based on topics. Most relevant to our work is the fact that the topic structures discovered from text using topic models can also be used to organize text. However, such a structure has various limitations in supporting user navigation~\cite{HoT} (e.g., only the major topics can be covered well). The hyperedges in HoT can cover all kinds of semantic links and cover them at the text segment level. As such, they are more useful than topic models for supporting navigation. 

{\bf Text clustering:} HoT construction can also be related to both text clustering~\cite{aggarwal2012survey,ahmed2022short} and segmentation~\cite{pak2018text}. While both of these tasks are focused on grouping text collections, traditional clustering approaches partition objects into non-overlapping sets. This makes traditional text clustering approaches ill-suited for HoT construction since a HoT with non-overlapping hyperedges would only enable within cluster navigation. By comparison, HoT allows for overlapping hyperedges meaning nodes can be a member of multiple sets. This enables broader connectivity making it a more suitable framework for supporting browsing. 

{\bf Semantic Matching, Hyperlink Generation, and Semantic Labeling:} Much work has been done on semantic matching of text~\cite{li2014semantic,lin2022pretrained}, mostly for applications such as information retrieval, question answering, text clustering, and plagiarism detection~\cite{Plagiarism19}. Some recent methods are based on LLMs~\cite{NEURIPS2023_f5708199}. Some work (e.g.,~\cite{smeaton1995experiments}) has also used similarity measure to create links to convert linear text into hypertext. Indeed, the automatic generation of hypertext and hyperlinks has also been studied extensively ~\cite{Yang2005ATM, Wilkinson1999AutomaticLG}. While the point-to-point linking of documents achieved by both semantic text matching and automatic hyperlink generation can indeed facilitate browsing, they do not enable topical browsing. Since HoT can enable both point-to-point browsing (via size 2 hyperedges) and topical browsing, it is a more general framework than older paradigms. 

Some recent work has also explored using LLMs for semantic tagging of text data (e.g.,~\cite{chen-etal-2021-hierarchy,Wan2024}, where a semantic tag from a hierarchy may be assigned to a text document, essentially enabling hierarchical text classification. Like text clustering, the objective of this task is similar to that of HoT construction. In fact, semantic labeling is even closer to HoT in that the groups have natural language labels. However, like text clustering, semantic labeling / tagging generally focuses on sorting texts into disjoint groups. While we can take inspiration from their approaches, the computational tasks of HoT and semantic tagging are ultimately different due to the focus of HoT construction on browsing requiring that nodes be members of many groups.  

{\bf Large Language Models (LLMs):} Recent years have seen explosive growth of work on applications of LLMs in many natural language processing~\cite{min2023recent} and information retrieval tasks~\cite{zhai2024large}, including, e.g., text summarization~\cite{zhang2024systematic}, information extraction~\cite{xu2023large} and question answering~\cite{tan2023can}. Our work introduces a new application of LLMs for HoT construction, which has not been explored before and directly supports semantic browsing in a large document space.

%% file: sections/HoT_Construction.tex
\section{Hypergraph of Text}
As discussed earlier, while users use both querying and browsing in information seeking, 
the support for querying is far more mature than that for browsing. While much progress has been made in search engine technology, we do not yet have any similar general technology for turning 
an arbitrary collection of scattered content into a well-organized linked content for general browsing support. One major barrier here is the difficulty in formalization of the browsing as a computational problem; without a well-defined computational problem for browsing, it is hard to study or evaluate any algorithms. 

A recent study has suggested that a browsable collection can be formally represented as a Hypergraph of Text (HoT)~\cite{HoT}. This work sheds some light on how to remove this long-standing barrier since we can now formally define the problem of supporting browsing as a computational problem, where the input is any collection of documents and the output is a HoT with each document as a node. The HoT can thus serve as a general mathematical structure for organizing scattered content. A significant benefit of framing the browsing problem in this way is that once a HoT is constructed, it can be visually displayed to a user using a general system such as the TINK system~\cite{TINK25} for browsing. This means that the same browsing system can be used to support browsing any HoT. Therefore, the whole problem of how to provide general support for browsing can now be reduced to one of how to construct a HoT for any arbitrary collection of documents. A main goal of this paper is to develop algorithms for constructing a HoT and, crucially, develop a method to evaluate a constructed HoT. 

As background, we first provide a brief introduction to the Hypergraph of Text (HoT) structure as proposed in~\cite{HoT}. We start with their definition: 
\begin{definition}[Hypergraph of Text] ~ \\
    A Hypergraph of Text is a hypergraph where the nodes and hyperedges have textual elements. Formally, we can define it as a hypergraph $H = (V,E, X_V,X_E)$ where
    \begin{enumerate}
        \item V is a set of nodes
        \item E is a set of hyperedges 
        \item $X_V = \{x_v\}_{v\in V}$ is a set of texts such that $\forall x_v \in X_V, x_v \in Z^{L_v}$
        \item $X_E = \{x_e\}_{e\in E}$ is a set of texts such that $\forall x_e \in X_E, x_e \in Z^{L_e}$
    \end{enumerate}
    Where $Z$ is the vocabulary and $L_v$ and $L_e$ are the length of the text sequence for node v and hyperedge e respectively. 
\end{definition}
This definition, which can be seen as a generalization of a textual graph, provides a foundation for navigation and various types of analysis when applied to a text collection. 

The skeptical reader may be wondering: what is gained from generalizing beyond graphs? To answer this question, we can return to foundational works in browsing. In the seminal work articulating the berrypicking model of information systems \cite{bates1989design}, Bates articulates six techniques for browsing:
\begin{enumerate}
    \item Footnote chasing
    \item Citation searching
    \item Journal run
    \item Area scanning
    \item Subject searching
    \item Author searching
\end{enumerate}
We make the following observation on Bates' six methods: of these six methods, only footnote chasing and citation searching are point-to-point. In the other four, the user is interested in groups of documents (either authorial groups as in author searching or topical groups in the rest). Therefore, to best support all modes of browsing we need a structure that can support both point-to-point connections (as a standard graph would) and n-ary connections. A hypergraph is precisely this structure. 

%% file: sections/methods.tex
\section{Methods}
In this section we discuss three approaches to HoT construction and their corresponding implementations. The first approach, ``All-Words'' treats all words in a document as a hyperedge and then weights and prunes hyperedges using TF-IDF scores. The second method, ``LLM-Doc'' and ``LLM-Sentence'' uses LLMs to extract topics at both sentence and document levels. Finally, our two-step similarity based approach combines semantic similarity measurements with LLM-guided topic extraction attempting to address weaknesses of the previous methods. 

In our discussion of these methods, we will mention the use of LLM and the use of a sentence transformer. In our implementations we use \texttt{llama3-8b-instruct} \cite{llama3} as our LLM. For sentence transformers, we utilize the sentence transformers package \cite{reimers-2019-sentence-bert} with \texttt{All-MiniLM-L6-v2} as our particular model \cite{wang2020minilmdeepselfattentiondistillation}. Full code for each of these methods, including LLM prompts, will be made available on GitHub. 
\subsection{All-Words}
\label{sec:allwords}
The core idea behind this approach is to treat ``All words'' shared between documents as a potential hyperedge. Naturally, including all words would be superfluous for navigation or analysis. Therefore, we only include those words which have a high average TF-IDF score (i.e. the top n \% of hyperedges as measured by average TF-IDF). 

More specifically, we can describe our approach as follows: First, create a hyperedge for each word in the document collection $\mathcal{C}$. For a given hyperedge $e$ which represents (and has its text $x_e$ set to) word $W_e$, a document $d_i$ is a member of a hyperedge $e$ if the term $W_e \in d_i$.  We then give a score to each hyperedge $e$ according to the following formula:
\[\textsc{Score}(e) = \frac{1}{|e|} \sum_{d \in e} \text{TF-IDF}(W_e, d)\]
Finally, we sort the hyperedges by $\textsc{Score}(e)$ and prune the HoT to include only the top $n\%$ of hyperedges. 
\subsection{LLM-Doc and LLM-Sentence}
\label{sec:llmsld}
Prior work indicated that LLMs could be used to create a HoT \cite{HoT}. In particular, they split each document into its sentences and prompt an LLM to extract the topics of each sentence and output them as a list. The prior work noted that this approach created two orders of magnitude more topics than documents. In order to avoid this, we propose an additional LLM based HoT construction method. The method is largely similar to the prior method, except topics are extracted at document level instead of sentence level.
\subsection{Two-Step Similarity}
\label{sec:twostepsld}
Initial investigation of prior methods indicated that while LLM showed promise, it was not without issue. In particular, the sentence level approach trended towards creating too many hyperedges and the document level approach trended towards too few. In order to focus the LLM on extracting topics with both fine-grained detail and relevance to the collection, we propose first picking sentence pairs and then using an LLM to extract the semantic links between these predetermined sentences. 

To ensure we only consider the most informative sentences, we use a two-step filtering process combining TF-IDF scores and embedding cosine similarity. We compute the average TF-IDF score for each sentence, allowing us to give a single number representing the importance of a sentence. The average TF-IDF score is as follows, where $v_i$ is the cached TF-IDF vector representation of sentence $s_i$:
\[score(s_i)=\frac{1}{|s_i|}\sum_j v_{i,j}\]
We compute the above score for each sentence in each document. We then sort the scores per document and retain only the top k sentences per document. 

Following the filtering step, we next want to utilize the sentence embeddings to find the most similar sentences for each sentence. While this approach is already made faster due to the prior filtering step, we use the GPU version of FAISS \cite{johnson2019billion} to ensure efficient execution. Finally, the top-k sentence pairs are chosen according to the following logic: First attempt to pick the highest cosine similarity sentence pairs without repeating sentences or documents, if this is not possible fill in the remaining sentence pairs allowing for repeats. This approach was chosen instead of simply choosing the top k pairs by cosine similarity in order to promote a diversity in connections. 

For each pair of sentences, we use LLM to extract the common topic. Each common topic becomes a hyperedge featuring the documents from the pair. Any subsequent documents also found to share this topic are accordingly added to the hyperedge. Unlike the "All-Words" method, this approach is quite usable without any additional pruning. However, the method does produce a lot of semantic links that are based on topics shared only by two documents. While it is well-defined for a hyperedge to have two members, one might also be interested in pruning these size-two hyperedges to focus on more substantial topical groups. We explore both a pruned version and a non-pruned version in our results. 

%% file: sections/effort_ratio.tex
\section{Evaluating HoT Construction}
\label{sec:eval}
How to evaluate HoT construction is a challenging problem. Indeed, there are no prior quantitative metrics that indicate the quality of a HoT for the purpose of browsing. This challenge is further complicated by the fact that there are many potentially useful ways to create hyperedges. To mitigate this problem, we introduce a general methodology for quantitatively evaluating the quality of a HoT's structure.

The core idea in this evaluation methodology is based on one of the goals of HoT: enabling navigation. If we have prior information indicating which documents are relevant to each other, we can compare the distances between these known relevant nodes and the distances between random nodes. The intuition is that if in a HoT, the average distance (as measured in hops on the hypergraph) of known relevant nodes to each other is 2 while the average distance of random nodes to each other is 4 then the produced hyperedges are useful because they allow for lower effort traversal between relevant nodes. Suppose we have a family of subsets $R = \{r_1, r_2, \dots\} \subseteq \mathcal{P}(\mathcal{C})$ where $\mathcal{C}$ is the set of documents and each $r \in R$ is a set of documents that are relevant to each other (for example, $R = \{\{a,b,c\}, \{a,c,e,f\}, \dots\}$; note that a document may appear in multiple sets of $R$). Suppose also that we generate a HoT $H$ from $\mathcal{C}$. The average relevant distance on $H$ is computed as the average, across every set $r \in R$, of the mean pairwise distance between documents in $r$. We call this measure DRel:
\[
\text{DRel} = \frac{1}{|R|} \sum_{r \in R} \frac{1}{|r|(|r| - 1)} \sum_{\substack{d_i, d_j \in r \\ i \neq j}} d_H(d_i, d_j)
\]
Note that if a pair $(d_i, d_j)$ appears in multiple $r \in R$, it contributes once per containing set; this weights pairs by the number of relevance groups they co-occur in, which we view as desirable.
Similarly, if we define a family of subsets $R' = \{r_1', r_2', \dots\} \subseteq \mathcal{P}(\mathcal{C})$ where each $r' \in R'$ is a randomly sampled set of documents, we define the average random distance as follows:
\[
\text{DRand} = \frac{1}{|R'|} \sum_{r \in R'} \frac{1}{|r|(|r| - 1)} \sum_{\substack{d_i, d_j \in r \\ i \neq j}} d_H(d_i, d_j)
\]
In both of these definitions, we take $d_H(d_i, d_j)$ to be the shortest path distance between nodes $d_i$ and $d_j$ on the generated hypergraph $H$. We treat the hypergraph as inducing an undirected graph in which two nodes $u, v \in V$ are adjacent iff there exists a hyperedge $e \in E$ with $\{u, v\} \subseteq e$; $d_H(u, v)$ is the length of the shortest path in this induced graph (and is undefined when $u$ and $v$ lie in different connected components, which we handle separately via RDP, defined below).

We can take the ratio of these measures $\text{DRel}/\text{DRand}$ to be the ``effort ratio'':
\[\text{ER} = \frac{\text{DRel}}{\text{DRand}}\]

An effort ratio of 1 indicates no benefit over random navigation; values below 1 indicate that relevant documents are closer than random documents; values above 1 would indicate that relevant documents are paradoxically farther from each other than the random set. Intuitively, if $\text{DRel} = 2$ and $\text{DRand} = 5$ then $\text{ER} = 0.4$, meaning that navigating between relevant nodes requires on average 40\% as many hops as navigating between random pairs of nodes.

This effort ratio will be the primary metric we use for evaluation. It is worth reiterating that this evaluation methodology and these metrics are quite general for evaluating HoT construction. While we choose a particular class of dataset in our experiments, there are many ways in which one can find documents with known relevance. A few examples include web logs, survey papers, and question answering datasets.

\subsection{Properties of Effort Ratio}
The effort ratio has several key features that make it well-suited for evaluating HoT construction. In this section, we spend some time walking through some of these properties. We begin with a few useful definitions.
\begin{definition}[Relevance-Aligned Hyperedge]~\\
    Let $H = (V, E, X_V, X_E)$ be a HoT. A hyperedge $e \in E$ is $\alpha$-relevance-aligned if:
    \[\frac{\left|\{\{d_i, d_j\} : d_i, d_j \in e,\ i \neq j,\ \exists\, r \in R \text{ s.t. } d_i, d_j \in r \}\right|}{\binom{|e|}{2}} \geq \alpha\]
\end{definition}
\begin{definition}[Non-Relevance-Aligned Hyperedge]~\\
    Let $H = (V, E, X_V, X_E)$ be a HoT. A hyperedge $e \in E$ is $\beta$-non-relevance-aligned if:
    \[\frac{\left|\{\{d_i, d_j\} : d_i, d_j \in e,\ i \neq j,\ \exists\, r \in R \text{ s.t. } d_i, d_j \in r \}\right|}{\binom{|e|}{2}} < \beta\]
\end{definition}
In other words, a hyperedge is $\alpha$-relevance-aligned if at least an $\alpha$ fraction of the (unordered) pairs of documents in the hyperedge are relevant to each other. Likewise a hyperedge is $\beta$-non-relevance-aligned if fewer than a $\beta$ fraction of pairs of documents are relevant to each other. Note that both numerator and denominator count unordered pairs, so the ratio lies in $[0, 1]$.

\begin{definition}[Saturation Measures]
    For a hypergraph $H$, we define the following two measures of saturation:
    \[\sigma_{\text{rel}} = \frac{\sum_{r \in R} \sum_{\substack{d_i, d_j \in r \\ i \neq j}}[d_H(d_i, d_j) = 1]}{\sum_{r \in R} \sum_{\substack{d_i, d_j \in r \\ i \neq j}} 1}\]
    \[\sigma_{\text{rand}} = \frac{\sum_{r \in R'} \sum_{\substack{d_i, d_j \in r \\ i \neq j}}[d_H(d_i, d_j) = 1]}{\sum_{r \in R'} \sum_{\substack{d_i, d_j \in r \\ i \neq j}} 1}\]
    where $[\cdot]$ denotes the Iverson bracket and $R$ and $R'$ are the same relevance and random sets as defined above.
\end{definition}
$\sigma_{\text{rel}}$ and $\sigma_{\text{rand}}$ can be understood as the fraction of pairs of documents that are one hop away for relevant and random documents respectively.

With these definitions in place, we now move on to characterizing effort ratio under different conditions. We start with investigating what happens when we add poorly-aligned edges to the hypergraph.

\subsubsection{Impact of non-relevance-aligned hyperedges}
\begin{proposition}[Informal]\label{prop:non-rel}
Let $H$ be a hypergraph with $\text{ER}(H) < 1$, and let $e$ be a candidate hyperedge that is $\beta$-non-relevance-aligned for some $\beta < 0.5$. Assume that (i) all pairs $(d_i, d_j) \subseteq e$ have finite distance in $H$ prior to adding $e$, and (ii) the expected shortcut (reduction in shortest-path distance) per non-relevant pair in $e$ is at least as large as the expected shortcut per relevant pair in $e$. Then adding $e$ to $H$ (weakly) increases $\text{ER}$.
\end{proposition}
The intuition behind Proposition~\ref{prop:non-rel} is that a $\beta$-non-relevance-aligned edge with $\beta < 0.5$ shortcuts more non-relevant pairs than relevant ones, so under the stated symmetry assumption on pairwise shortcut magnitudes it decreases $\text{DRand}$ by more than it decreases $\text{DRel}$, pushing $\text{ER}$ up toward 1. A full proof tracks, for each pair $(d_i, d_j) \subseteq e$, the change in $d_H(d_i, d_j)$ (which is at most $1 - d_H(d_i, d_j)$ since $e$ creates a single-hop path) and sums these changes over the relevant- and non-relevant-pair partitions.

However, Proposition~\ref{prop:non-rel} can fail when the assumptions do not hold. The most extreme case is $\sigma_{\text{rand}} = 1$ and $\sigma_{\text{rel}} < 1$. In this case, adding an edge $e$ that is $\beta$-non-relevance-aligned for $\beta < 0.5$ but is $\alpha$-relevance-aligned for some $\alpha > 0$ (i.e.\ any edge containing at least one relevant pair) can actually decrease the effort ratio. Consider: if $\sigma_{\text{rand}} = 1$ then $\text{DRand} = 1$, and adding further edges cannot change that. However, if $\sigma_{\text{rel}} < 1$ then $\text{DRel} > 1$, and shortcutting any of the remaining non-connected relevant pairs decreases $\text{DRel}$. So the resulting decrease in effort ratio comes from moving it from above 1 down toward 1.

The key insight here is that, except in pathological cases where the hypergraph structure is highly imbalanced and random nodes are more closely connected than relevant nodes, the effort ratio penalizes adding hyperedges that connect more irrelevant documents than relevant ones. This is exactly the type of behavior we would want in a navigation metric.

Another desirable property of a navigation metric is the ability to penalize overload. That is to say, we do not want to give a good score to a method that achieves good connections by simply spamming edges. We want a metric that has some regularizing effect on the total number of edges in the hypergraph. This leads us to our next result.

\subsubsection{Moderation of hypergraph size}
\begin{proposition}[Informal]\label{prop:saturation}
Let $H_0$ be a finite hypergraph over $\mathcal{C}$, and let $\{e_1, e_2, \dots\}$ be a sequence of hyperedges, each $\alpha$-relevance-aligned for some fixed $\alpha \in (0, 1]$, where for every $n$ the pairs contained in $e_n$ are drawn from a distribution over $\binom{\mathcal{C}}{2}$ that assigns positive probability to every pair. Let $H_n$ be the hypergraph obtained by adding $e_1, \dots, e_n$ to $H_0$, and assume $H_n$ is connected for all sufficiently large $n$. Then, almost surely,
\[\lim_{n\rightarrow\infty}\frac{\text{DRel}_n}{\text{DRand}_n} = 1.\]
\end{proposition}
The intuition is a coupon-collector-style argument: even ``good'' edges that are $\alpha$-relevance-aligned for $\alpha$ close to 1 eventually saturate the hypergraph. Because every pair has positive sampling probability, every pair (relevant and random alike) is eventually one hop apart, so both $\text{DRel}_n$ and $\text{DRand}_n$ tend to 1 and their ratio tends to 1. The connectedness assumption ensures the distances are finite at each step; under the positive-probability assumption this occurs almost surely for some finite $n$.

\subsubsection{Limitations of Effort Ratio}
While effort ratio allows quantitative evaluation of HoT, it does have two key limitations. First, it is not well defined if the hypergraph has disconnected components. This is the direct result of the metric being based on distances between nodes. As a result, effort ratio should always be considered alongside the extent to which items are disconnected. For this reason, we propose another metric: Relevant Disconnect Proportion (RDP). This metric is simply the proportion of pairs of relevant documents which are disconnected from each other. Taken together with effort ratio, RDP allows us to know if the HoT construction method is essentially cheating the metric via selection bias.

The second key limitation of effort ratio is that it does not measure the coherence of the hyperedge with the text describing the hyperedge. While it does implicitly measure the semantic coherence of the hyperedge itself, the hyperedge text could be anything. For example, if one had a HoT $H$ and created a new HoT $H'$ by changing the text associated with every hyperedge to be a random word, $H$ and $H'$ would have the same effort ratio. This indicates that it would likely be useful to also evaluate the text of the hyperedges in HoTs in addition to the hypergraph structure. We leave this for future work.

%% file: sections/results.tex
\section{Experimental Results}
 For our experiments we use the metrics proposed in section \ref{sec:eval} on a MultiHop-RAG QA data set \cite{tang2024multihopragbenchmarkingretrievalaugmentedgeneration}. This dataset consists of 609 articles and 2556 multi-hop queries distributed across 2 to 4 documents. While the MultiHop-RAG paper does not indicate any topic distribution, manual inspection shows topics including current events, financial markets and stocks, sporting events, and video games. For the purposes of our experiments, we take each query associated with documents to be a set of documents relevant to each other. This means that our set $R$ (as described in section \ref{sec:eval}) is a set of sets size 2556 where each member set is size 2 to 4.

\begin{table}[!ht]
    \centering
    \small
        \caption{Comparative analysis of HoT construction methods. Lower effort ratio indicates better navigation support. Random Hypergraphs of various sizes are provided as a sanity check. RDP = Relevant Disconnect Proportion}
    \begin{tabular}{@{}lccc@{}}
    \toprule
        \textbf{Method} & \textbf{Effort Ratio} & \textbf{Number of} & \textbf{RDP} \\
        & & \textbf{Hyperedges} &  \\
    \midrule
        LLM-doc & 0.362 & 2,155 & 0.385 \\
        LLM-sentence & 0.919 & 21,295 & 0.000 \\
        \midrule
        Two-step LLM & 0.596 & 5,311 & 0.014 \\
        Two-step LLM (pruned) & 0.598 & 2,084 & 0.015 \\
        \midrule
        All-Words top 1\% & 0.523 & 232 & 0.199 \\
        All-Words top 5\% & 0.582 & 1,160 & 0.000 \\
        \midrule
        Random HG (200) & 0.989 & 200 & 0.037 \\
        Random HG (800) & 1.002 & 800 & 0.000 \\
        Random HG (1600) & 0.991 & 1,600 & 0.000 \\
    \bottomrule
    \end{tabular}

    \label{table:effort-results}
\end{table}

Our experiments revealed interesting differences in performance across the various HoT construction methods. The results are summarized in table \ref{table:effort-results}. Before getting deeper into the results, it is first worth noting the relative compute requirements of each method. The two `All-Words' methods are fairly lightweight and do not require any machine-learning-capable hardware. The LLM-based methods are much more computationally demanding requiring significant LLM usage. The two-step LLM methods also require LLM usage, but significantly less due to the filtering step. 

The LLM category of approaches displayed the most variance with the sentence level approach and document level approaches getting the highest (non-random) and lowest effort ratios respectively. This is, perhaps, unsurprising in that these two methods also show the largest discrepancy in number of hyperedges with the sentence level approach being an order of magnitude larger than the document level approach. While the document level approach achieved the lowest effort ratio of any of the proposed methods, it is important to note that almost 40\% of relevant document pairs were disconnected. Since disconnected documents cannot be counted in the average distances as they'd turn the average distance to infinity, this method having the lowest effort ratio could be indicating that it was only able to make the ``easy'' connections. Methods with lower disconnect rates must account for these 'hard' connections, which are typically farther apart, thus incurring a penalty to their average relevant distance.

In comparison, both the ``All-Words'' and ``two-step'' methods achieved results with effort ratios in the .50 to .60 range. Both `All-Words' methods perform well despite their fewer number of hyperedges. The caveat to this is that the semantic links created by the `All-Words' are single word topics. This is a quality limitation for their semantic links. For instance, unless a person is easily recognizable by a single name (e.g. Plato) then this method cannot capture topics related to that person as well. The ``two-step'' method does not have this downside. Since the ``two-step'' method takes considerably longer to run than ``All-Words'', the best method likely depends on one's limitations and use cases as each method has benefits and drawbacks.  For budget-constrained scenarios or scenarios where topic nuance are less important, All-Words at top-5\% is within 0.02 of Two-Step at ~1/5 the hyperedge count and no LLM calls.

%% file: sections/conclusion.tex
\section{Limitations}
As an initial study of HoT construction, we only study HoT Construction on one dataset. In the future, it is important to further verify our findings by experimenting with additional datasets by using the proposed simulation evaluation strategy. Once a HoT is constructed, many algorithms (e.g., random walk, path discovery, outlier discovery) can be applied to post-process HoT to reveal many interesting latent semantic patterns in the text data (e.g., controversy analysis, comparative analysis, or association analysis). Additionally, while the quantitative evaluation enabled us to compare multiple HoT construction approaches, the actual utility of HoT construction from a user's perspective still needs further evaluation using user studies. 

\section{Conclusion and Future Work}
In this paper, we study how to construct a Hypergraph of Text (HoT) to semantically organize an arbitrary text collection for supporting navigation. We study three approaches to performing HoT construction, including approaches that leverage LLMs. To quantitatively evaluate browsing, we propose a novel evaluation metric: effort ratio. Our experiment results show that while various methods are able to produce potentially useful links, each has its own benefits and drawbacks. With HoT as a foundation, our work makes a step toward developing a general technology to support a user in browsing an arbitrary collection of documents, which has widespread applications. An important next step is to further validate these metrics and algorithms using real user studies. Another promising future direction is to apply the proposed algorithms to specific application scenarios such as constructing a HoT to organize search results, research paper collections, organize any shared digital libraries created by a group of people (e.g., student essays in a class), or organize any personal folders of content.  